\documentclass[prd,nofootinbib]{revtex4}

\usepackage{amsmath,amsfonts,
amssymb,graphicx}

\newcommand{\npartial}{{\not\!\partial}}
\newcommand{\nA}{{\not\!\!A}}
\newcommand{\nk}{{\not\!k}}

\begin{document}
\title{Pair production with neutrinos and high-intensity laser fields}%
\date{\today}%
\author{Todd M. Tinsley}%
\affiliation{Center for Particle Physics, University of Texas at
Austin, Austin, Texas 78712, USA}%
\email{ttinsley@physics.utexas.edu}

\begin{abstract}
We present a detailed calculation of the electron-positron
production rate using neutrinos in an intense laser field.  The
computation is done for the process $\nu \to \nu e \bar{e}$ via
the neutral channel and within the framework of the Standard
Model.  The production rates are tabulated for a range of incoming
neutrino energies in an electromagnetic field similar to what can
be attained by current high-intensity lasers.
\end{abstract}

\maketitle

\section{Introduction\label{sec:intro}}
The introduction of a background electromagnetic field can have
very important consequences for neutrino physics. While the
Standard Model neutrino does not itself couple to an
electromagnetic field, the effects of such a field on neutrinos
are made manifest through their interactions with charged
particles, real or virtual.  Specifically, many investigations
have calculated the effects of a constant magnetic field on
neutrino propagation and interactions.  Recent areas of interest
for such calculations include: the neutrino self-energy
\cite{Erdas:1990gy,Elizalde:2004mw}, the $\gamma$-$\nu$-$\gamma$
vertex \cite{Gvozdev:1997mc}, nucleon interactions
\cite{Baiko:1998jq,Bandyopadhyay:1998qs,Leinson:2001ei},
neutrino-photon scattering
\cite{Dicus:2000cz,Shaisultanov:1997bc,Dicus:1997rw,%
Dicus:1998uv,Abbasabadi:1998ba,Dicus:2000tk,Dicus:2000uw},
and neutrino-electron interactions \cite{Hardy:2000gg}. A very
nice review of these examples and their astrophysical effects has
been prepared by Bhattacharya and Pal \cite{Bhattacharya:2002aj}.

In this article we examine electron-positron pair production
through the process $\nu \to \nu e \bar{e}$. Though normally
forbidden, the electromagnetic field alters the final states of
the electron-positron pair and frees the interaction to take
place. Such a process could have very significant astrophysical
and cosmological ramifications. Gvozdev \textit{et al.}\ have
proposed that this interaction could be important in the analysis
of supernovas and neutron star coalescence, and they have offered
its role in magnetars as a possible explanation for gamma ray
bursts \cite{Gvozdev:1997bs}.  Such effects are important because
of the magnetic field strengths associated with these phenomena.
For example, neutron stars and supernovae can have fields in
excess of $10^{12}\,\mathrm{G}$ \cite{Raffelt:1996wa}, while a
magnetar's field is of order $\lesssim 10^{15}\,G$
\cite{Thompson:1996pe}.

We, however, turn our focus to examining the production of pairs
with neutrinos within more terrestrial fields. We investigate the
feasibility of detecting $\nu \to \nu e \bar{e}$ events using
ultrashort pulsed lasers. Though not approaching the astronomical
magnetic fields, today's femtosecond terawatt lasers can produce
field strengths on the order of $\mathrm{E} \sim 10^{11}\,
\mathrm{V/cm}$ ($B \sim 10^{9}\,\mathrm{G}$) \cite{Ditmire}. We
begin by presenting a review of the derivation of the Volkov field
operator solution for the electron(positron) in an circularly
polarized electromagnetic field (Section~\ref{sec:derive}). Next,
we apply this Volkov solution to the calculation of the production
rate for the process $\nu \to \nu e \bar{e}$
(Section~\ref{sec:prodrate}).  Lastly, we provide a tabulation of
production rates for various incoming neutrino energies in the
laser field and estimate the likelihood of detecting such a
process (Section~\ref{sec:results}).

\section{Derivation of Field\label{sec:derive}}
As was pointed out in Section~\ref{sec:intro}, any effect an
electromagnetic field has on Standard Model neutrino physics is
due to interaction with charged particles in the interaction.  For
our process we incorporate the laser's intense field into the
problem by using the field operator solution of the electron in
the presence of an electromagnetic plane wave.  We begin by
solving the Dirac equation for such a case
\begin{equation}\label{eq:Dirac}
\left( \imath \hbar \npartial  - e \nA - m_{\mathrm{e}}c\right)
\psi(x) = 0\,.
\end{equation}
We choose the electromagnetic field to be circularly polarized and
directed along the $z$-direction such that the 4-vector potential
$A(x)$ is
\begin{equation}\label{eq:A}
A(x) = \left( \begin{array}{c} 0 \\ a \cos k \cdot x \\ a \sin k
\cdot x \\ 0 \end{array} \right)
\end{equation}
where $a$ is the magnitude of the vector potential and $\hbar k$
is the 4-momentum directed along the $z$-direction.  The choice of
circularly polarized light as already been pointed out to be a
much simpler approach than assuming linearly or ecliptically
polarized light \cite{Denisov:68,Szymanowski:97}.

D. M. Volkov solved the Dirac equation for an electron in the
presence of an electromagnetic plane wave
\cite{Volkov:35,Volkov:37}.  More recent derivations of the
solution are outlined in \cite[$\S\ 40$]{Lifshitz:82} and
\cite{Szymanowski:97}.  We follow the example of Lifshitz,
\textit{et al.} \cite{Lifshitz:82} and find that the Volkov field
operator $\psi(x)$ is
\begin{widetext}
\begin{eqnarray}\label{eq:vfieldop}
\psi(x) &=& \int \frac{d^3 \vec{p}}{(2\pi)^3} \exp\left[
\frac{\imath ea}{p\cdot k} \left( p_{1} \sin k\cdot x - p_{2} \cos
k\cdot x\right)\right]
\sum_s \Biggl\{ \left( 1 + \frac{e \nk \nA}{2 p\cdot k}
\right) \exp\left[ -\frac{\imath}{\hbar} \left( p + \frac{e^2
a^2}{2 p\cdot k} k\right) \cdot x \right]
\frac{u^s(p)}{\sqrt{2 E_{\vec{p}}/c}}\ a^s_{\vec{p}} \nonumber\\
&&
+ \left( 1 - \frac{e \nk \nA}{2 p\cdot k} \right) \exp\left[
+\frac{\imath}{\hbar} \left( p + \frac{e^2 a^2}{2 p\cdot k}
k\right) \cdot x \right] \frac{v^s(p)}{\sqrt{2 E_{\vec{p}}/c}}\
{b^s_{\vec{p}}}^\dag \Biggr\}\,.
\end{eqnarray}
\end{widetext}
In equation~(\ref{eq:vfieldop}) we have adopted the notation of
\cite{Peskin:95} where the summation over $s$ is a sum over
possible spin states.  We integrate over the particle's momentum
$p$, and the quantities $p_{1}$ and $p_{2}$ are the components of
that momentum along the $x$ and $y$ directions, respectively. We
note that the momentum that is to be identified with the Volkov
state is not $p$.  Instead the Volkov momentum, denoted as $q$, is
defined by
\begin{equation}\label{eq:vmomentum}
q^\mu = p^\mu + \frac{e^2 a^2}{2 p\cdot k} k^\mu\,.
\end{equation}
Squaring the state's 4-momentum shows that the mass of the Volkov
electron has a dependence on the strength of the electromagnetic
field
\begin{equation}\label{eq:vmass}
{m_e^\ast}^2 = {m_e}^2 + e^2 a^2 /c^2\,.
\end{equation}

Note that in the absence of an electromagnetic field $a \to 0$ the
Volkov field operator, equation~(\ref{eq:vfieldop}), reduces to
the free-field solution
\begin{eqnarray}\label{eq:vfieldoplimit}
\lim_{a \to 0} \psi(x) &=& \int \frac{d^3 \vec{p}}{(2\pi)^3}
\sum_s \Biggl\{ e^{-\frac{\imath}{\hbar} p \cdot x}
\frac{u^s(p)}{\sqrt{2 E_{\vec{p}}/c}}\ a^s_{\vec{p}} +
e^{+\frac{\imath}{\hbar} p\cdot x} \frac{v^s(p)}{\sqrt{2
E_{\vec{p}}/c}}\ {b^s_{\vec{p}}}^\dag
\Biggr\}\,,\label{eq:ffieldop}
\end{eqnarray}
and the Volkov mass $m_e^\ast$ reduces to the electron mass $m_e$.

\section{The Rate of Production {}$\mathbf{\Gamma}$\label{sec:prodrate}}
A quantity that we wish calculate for this process is the rate of
production $\Gamma$.  Physically, the rate of production is the
probability per unit time for the neutrino to emit an
electron-positron pair in the presence of the laser field. We
begin by finding the probability $\mathcal{P}$ for the interaction
\begin{equation}\label{eq:prob1}
\mathcal{P} = \left( \prod_f \int \frac{d^3 \vec{p}_f}{(2\pi)^3}
\frac{1}{2 E_f/c} \right) \int \frac{d^3 \vec{p}_{\nu}}{(2\pi)^3}
\frac{1}{2E_{\nu}/c}\ \frac{1}{2} \sum_{s_\nu} \sum_{s_{e},
s_{\bar{e}}, s_{\nu'}} \bigl| \langle p_{\nu'}, s_{\nu'}; p_{e},
s_{e}; p_{\bar{e}}, s_{\bar{e}} | \hat{S} | p_{\nu}, s_{\nu}
\rangle \bigr|^2
\end{equation}
where we have summed over the spin states of the final neutrino
$\nu'$, electron $e$, and positron $\bar{e}$, and averaged over
the spin states of the incoming neutrino $\nu$.  The phase-space
integral over the final momentums $p_e$, $p_{\bar{e}}$, and
$p_{\nu'}$ has been simplified into the form
\begin{equation}\label{eq:phasespace}
\prod_f \int \frac{d^3 \vec{p}_f}{(2\pi)^3} \frac{1}{2 E_f/c} =
\int \frac{d^3 \vec{p}_{\nu'}}{(2\pi)^3} \frac{1}{2 E_{\nu'}/c}
\int \frac{d^3 \vec{p}_e}{(2\pi)^3} \frac{1}{2 E_e/c} \int
\frac{d^3 \vec{p}_{\bar{e}}}{(2\pi)^3} \frac{1}{2
E_{\bar{e}}/c}\,.
\end{equation}

The scattering operator $\hat{S}$ for our process is
\begin{eqnarray}\label{eq:Sop}
\hat{S} &=& -\frac{\imath}{\hbar^8}\ \frac{4\pi \alpha}{2^2 \cos^2
\theta_{\mathrm{W}} \sin^2\theta_{\mathrm{W}}} \int d^4x\
\overline{\psi}_e(x) \gamma^\mu \left(-\frac{1}{2}+2
\sin^2\theta_{\mathrm{W}} + \frac{1}{2}
\gamma^5\right) \psi_e(x) Z_\mu(x) \nonumber\\
&& \times \int d^4y\ \overline{\psi}_{\nu}(y) \gamma^\sigma
\left(\frac{1}{2} - \frac{1}{2} \gamma^5\right) \psi_{\nu}(y)
Z_\sigma(y)
\end{eqnarray}
where $\psi_e(x)$ is the Volkov field operator for the electron
(equation~(\ref{eq:vfieldop})), $\psi_{\nu}(y)$ is the free-field
operator for the neutrino, and $Z_\mu(x)$ is the field operator of
the $Z$ boson.  $\alpha$ is the fine-structure constant, and
$\theta_{\mathrm{W}}$ is the weak-mixing angle.

In the free-field case the integration over the space-time
variables results in a 4-dimensional $\delta$-function that
conserves energy and momentum.  Though this is true of our
integral over the space-time coordinate $y$ in
equation~(\ref{eq:Sop}), it is not true for the integral over $x$.
The Volkov fields $\psi_e(x)$ in the $\int d^4x$ integral have a
more complicated space-time dependence.

We begin by expanding the field operators and integrating out the
$y$ dependence.  The scattering matrix can be written as
\begin{widetext}
\begin{eqnarray}\label{eq:Smatrix2}
S &=&  \langle p_{\nu'}, s_{\nu'}; p_{e}, s_{e}; p_{\bar{e}},
s_{\bar{e}} | \hat{S} | p_{\nu}, s_{\nu} \rangle \nonumber \\
S &=& -\frac{\imath}{\hbar^4}\ \frac{4\pi \alpha}{2^2 \cos^2
\theta_{\mathrm{W}} \sin^2\theta_{\mathrm{W}}} \left\{
\overline{u}_{s_{\nu'}}(p_{\nu'}) \gamma^\sigma \left(\frac{1}{2}
- \frac{1}{2} \gamma^5\right) u_{s_\nu}(p_{\nu})\right\}
\frac{g_{\mu\sigma} - (p_{\nu}-p_{\nu'})_\mu
(p_{\nu}-p_{\nu'})_\sigma/(m_Z c)^2}{(p_{\nu}-p_{\nu'})^2 - (m_Z
c)^2 + \imath \Gamma_Z m_Z c^2} \nonumber\\
&&\times \int d^4x\ \left\{ \overline{u}_{s_e}(p_e) \left(1 +
\frac{e \nA \nk}{2 p_e \cdot k}\right) \gamma^\mu
\left(-\frac{1}{2}+2 \sin^2\theta_{\mathrm{W}} + \frac{1}{2}
\gamma^5\right) \left(1 - \frac{e \nk \nA}{2 p_{\bar{e}} \cdot
k}\right)
\overline{u}_{s_{\bar{e}}}(p_{\bar{e}})\right\}\nonumber\\
&& \times  \exp\left[\frac{\imath}{\hbar} (q_e + q_{\bar{e}} +
p_{\nu'} - p_\nu) \cdot x\right] \exp\left[ -\left( \frac{ea}{p_e
\cdot k} {p_e}_1 - \frac{ea}{p_{\bar{e}} \cdot k} {p_{\bar{e}}}_1
\right) \sin(k \cdot x) + \left(\frac{ea}{p_e \cdot k} {p_e}_2 -
\frac{ea}{p_{\bar{e}} \cdot k} {p_{\bar{e}}}_2 \right) \cos(k
\cdot x)\right]. \nonumber\\
&&
\end{eqnarray}
\end{widetext}

To fashion equation~(\ref{eq:Smatrix2}) into an analytically
integrable form, we employ Hansen's definition of the Bessel
function \cite{Weisstein}
\begin{equation}\label{eq:HdefBess}
e^{(z/2)(t-1/t)} = \sum_{n=-\infty}^{\infty} J_n(z) t^n\,.
\end{equation}
We can recast the left-hand side of equation~(\ref{eq:HdefBess})
in a fashion more applicable to the exponentials found in
equation~(\ref{eq:Smatrix2})
\begin{equation}\label{eq:exptobess}
e^{-\imath \zeta \sin(k\cdot x - \delta)} =
\sum_{n=-\infty}^{\infty} e^{-\imath n k \cdot x} e^{\imath n
\delta} J_n(\zeta)\,,
\end{equation}
where
\begin{subequations}\label{eq:zddef}
\begin{eqnarray}
\zeta \cos\delta &=& \left( \frac{ea}{p_e \cdot k} {p_e}_1 -
\frac{ea}{p_{\bar{e}} \cdot k} {p_{\bar{e}}}_1 \right)\,,\
\textrm{and} \label{eq:zcosd} \\
\zeta \sin\delta &=& \left( \frac{ea}{p_e \cdot k} {p_e}_2 -
\frac{ea}{p_{\bar{e}} \cdot k} {p_{\bar{e}}}_2 \right)\,.
\label{eq:zsind}
\end{eqnarray}
\end{subequations}

Since the vector potential $A$ in the $S$-matrix has components
that go as $\sin(k\cdot x)$ and $\cos(k\cdot x)$, we will also
need to perform a similar expansion for terms like
$$
\left\{\begin{array}{c}
\sin(k\cdot x)\\
\cos(k\cdot x)\\
\sin^2(k\cdot x)\\
\sin(k\cdot x)\cos(k\cdot x)\\
\cos^2(k\cdot x)\end{array}\right\} \times e^{-\imath \zeta
\sin(k\cdot x - \delta)}\,.
$$
All of these terms, when expanded, result in forms similar to
equation~(\ref{eq:exptobess}).  The space-time dependence is
factored out into an exponential that goes as $e^{-\imath n k
\cdot x}$. As an example,
\begin{equation}\label{eq:sinexpexpan}
\sin(k \cdot x) e^{-\imath \zeta \sin(k\cdot x - \delta)} =
\sum_{n=-\infty}^{\infty} \frac{1}{2\imath} e^{-\imath n k\cdot x}
\left( e^{\imath(n+1) \delta} J_{n+1}(\zeta) - e^{\imath(n-1)
\delta} J_{n-1}(\zeta) \right)\,.
\end{equation}

After these expansions are made, the space-time integral in the
$S$-matrix reduces to the normal 4-dimensional $\delta$-function
\begin{eqnarray}
S &\propto & \int d^4x\ \exp\left[\frac{\imath}{\hbar} (q_e +
q_{\bar{e}}
+ p_{\nu'} - p_{\nu} - n k) \cdot x \right] \nonumber\\
S &\propto & \delta^4(q_e + q_{\bar{e}} + p_{\nu'} - p_{\nu} - n
k)\,. \label{eq:deltafunc}
\end{eqnarray}
Physically, equation~(\ref{eq:deltafunc}) suggests that it is the
Volkov states that are to be considered when conserving momentum
and energy.  Equation~(\ref{eq:deltafunc}) also suggests that we
interpret the $nk$ term as a number of photons that are pulled
from the laser field in order to drive this reaction.  These
photons are what allow this process to overcome the energy and
momentum constraints that originally prohibited this reaction in
the free-field case. In fact, the $\delta$-function can be used to
apply a constraint on the possible values for $n$
\begin{eqnarray}
{p_{\nu}}^\mu + n k^\mu &=& {q_e}^\mu + {q_{\bar{e}}}^\mu +
{p_{\nu'}}^\mu \nonumber\\
\left({p_{\nu}}^\mu + n k^\mu\right)^2 &=& \left({q_e}^\mu +
{q_{\bar{e}}}^\mu + {p_{\nu'}}^\mu\right)^2 \nonumber\\
2 n {p_{\nu}} \cdot k &\geq & 4 (m_e^\ast c)^2\,.
\label{eq:nconstraint}
\end{eqnarray}
In the proceeding calculation we will take the neutrino and laser
beams to be oppositely directed.  Therefore, in the massless limit
of the neutrino the constraint equation~(\ref{eq:nconstraint})
becomes
\begin{eqnarray}
4 n E_{\nu} E_\gamma &\geq & 4 (m_e^\ast c)^2 \nonumber\\
n &\geq &  \frac{(m_e^\ast c)^2}{E_{\nu} E_\gamma} \,,
\label{eq:nconstraint2}
\end{eqnarray}
where $E_\gamma$ is the energy carried per photon.  We note that
the reaction cannot emit photons ($n \nless 0$), nor can it
proceed without absorbing some number of photons ($n \neq 0$).

The $S$-matrix can now be rewritten as
\begin{equation}\label{eq:Smatrix3}
S = \imath (2\pi)^4 \sum_{n=-\infty}^{\infty} \delta^4(q_e +
q_{\bar{e}} + p_{\nu'} - p_{\nu} - n k)\ \mathcal{M}_n\,
\end{equation}
where we have absorbed the couplings, spinors, Bessel functions,
etc., into the scattering matrix $\mathcal{M}_n$.  For a
definition of $\mathcal{M}_n$ refer to the Appendix.

Of importance, however, is the square of the $S$-matrix.  We begin
by substituting the $S$-matrix (equation~(\ref{eq:Smatrix3})) into
the probability (equation~(\ref{eq:prob1}))
\begin{widetext}
\begin{eqnarray}\label{eq:prob2}
\mathcal{P} &=& \left( \prod_f \int \frac{d^3 \vec{p}_f}{(2\pi)^3}
\frac{1}{2 E_f/c} \right) \int \frac{d^3
\vec{p}_{\nu}}{(2\pi)^3} \frac{1}{2E_{\nu}/c} \nonumber\\
&&\times \frac{1}{2} \sum_{s_\nu} \sum_{s_{e}, s_{\bar{e}},
s_{\nu'}} \sum_{n,m=-\infty}^{\infty} (2\pi)^8 \delta^4(q_e +
q_{\bar{e}} + p_{\nu'} - p_{\nu} - n k)\ \delta^4(q_e +
q_{\bar{e}} + p_{\nu'} - p_{\nu} - m k)\ \mathcal{M}_m^\ast
\mathcal{M}_n\,.
\end{eqnarray}
\end{widetext}
By inspection, one can see that the two 4-dimensional
$\delta$-functions imply that for there to be any contribution to
the summation, either there is no incoming photon energy
($E_\gamma = 0$) or, more appropriately, that $m = n$. Therefore,
we can replace $\mathcal{M}_m^\ast \mathcal{M}_n$ with the square
of the norm of the scattering amplitude $\bigl|
\mathcal{M}_n\bigr|^2$ and eliminate the sum over $m$
\begin{equation}\label{eq:prob3}
\mathcal{P} = \left( \prod_{f} \int \frac{d^3
\vec{p}_{f}}{(2\pi)^3} \frac{1}{2E_{f}/c} \right) \int \frac{d^3
\vec{p}_{\nu}}{(2\pi)^3} \frac{1}{2E_{\nu}/c}
\sum_{n=-\infty}^{\infty} (2\pi)^8 \left(\delta^4(q_e +
q_{\bar{e}} + p_{\nu'} - p_{\nu} - n k)\right)^2\ \overline{\bigl|
\mathcal{M}_n\bigr|^2}\,.
\end{equation}
To simplify the result, we have defined the square of the
scattering matrix after summing and averaging over spins to be
\begin{equation}\label{eq:avgM2}
\overline{\bigl| \mathcal{M}_n\bigr|^2} = \frac{1}{2} \sum_{s_\nu}
\sum_{s_{e},s_{\bar{e}},s_{\nu'}} \bigl| \mathcal{M}_n\bigr|^2\,.
\end{equation}

Next, we can use three of the $\delta$-functions to eliminate the
integral over the momentum of the incoming neutrino
\begin{eqnarray*}
\mathcal{P} &=& \left( \prod_{f} \int \frac{d^3
\vec{p}_{f}}{(2\pi)^3} \frac{1}{2E_{f}/c} \right)
\frac{1}{2E_{\nu}/c} \\
&& \times \sum_{n=-\infty}^{\infty} (2\pi)^5 \delta^4(q_e +
q_{\bar{e}} + p_{\nu'} - p_{\nu} - n k) \delta\bigl((E_e +
E_{\bar{e}} + E_{\nu'} - E_{\nu} - n E)/c\bigr)\ \overline{\bigl|
\mathcal{M}_n\bigr|^2}\,,
\end{eqnarray*}
or
\begin{eqnarray}\label{eq:prob4}
\mathcal{P} &=& \left( \prod_{f} \int \frac{d^3
\vec{p}_{f}}{(2\pi)^3} \frac{1}{2E_{f}/c} \right)
\frac{1}{2E_{\nu}/c} \nonumber\\
&& \times \sum_{n=-\infty}^{\infty} (2\pi)^5 \delta^3(\vec{q}_e +
\vec{q}_{\bar{e}} + \vec{p}_{\nu'} - \vec{p}_{\nu} - n \vec{k})
\Bigl(\delta\bigl((E_e + E_{\bar{e}} + E_{\nu'} - E_{\nu} - n
E)/c\bigr)\Bigr)^2\ \overline{\bigl| \mathcal{M}_n\bigr|^2}\,.
\end{eqnarray}
This still leaves us with two identical energy-conserving
$\delta$-functions.  To simplify we follow the solution pointed
out by \cite{Szymanowski:97} and outlined in \cite{Sakurai:67}. We
write one of these $\delta$-functions as an integral over time
\begin{eqnarray*}
\mathcal{P} &\propto& \sum_{n=-\infty}^{\infty}
\Bigl(\delta\bigl((E_e + E_{\bar{e}} + E_{\nu'} - E_{\nu} - n
E)/c\bigr)\Bigr)^2 \\
\mathcal{P} &\propto& \sum_{n=-\infty}^{\infty} \delta\bigl((E_e +
E_{\bar{e}} + E_{\nu'} - E_{\nu} - n E)/c\bigr)
\frac{c}{2\pi\hbar}\ \lim_{T\to \infty} \int_{-T/2}^{T/2} dt\
e^{\frac{\imath}{\hbar} (E_e + E_{\bar{e}} + E_{\nu'} - E_{\nu} -
n E) t}\,,
\end{eqnarray*}
and use the other $\delta$-function to reduce the integrand to
unity
\begin{eqnarray}\label{eq:intovertime}
\mathcal{P} &\propto& \sum_{n=-\infty}^{\infty} \delta\bigl((E_e +
E_{\bar{e}} + E_{\nu'} - E_{\nu} - n E)/c\bigr)\
\frac{c}{2\pi\hbar}\ \lim_{T\to \infty} \int_{-T/2}^{T/2} dt
\nonumber\\
\mathcal{P} &\propto& \sum_{n=-\infty}^{\infty} \delta\bigl((E_e +
E_{\bar{e}} + E_{\nu'} - E_{\nu} - n E)/c\bigr)\
\frac{c}{2\pi\hbar}\ \lim_{T\to \infty} T \,.
\end{eqnarray}

By making the substitution of equation~(\ref{eq:intovertime}) into
the probability in equation~(\ref{eq:prob4})
$$
\mathcal{P} = \lim_{T\to \infty}\ \left( \prod_{f} \int \frac{d^3
\vec{p}_{f}}{(2\pi)^3} \frac{1}{2E_{f}/c} \right)
\frac{1}{2E_{\nu}/c}\ \frac{c}{2\pi\hbar}\ T
\sum_{n=-\infty}^{\infty} (2\pi)^5 \delta^4(q_e + q_{\bar{e}} +
p_{\nu'} - p_{\nu} - n k)\ \overline{\bigl|
\mathcal{M}_n\bigr|^2}\,,
$$
we can solve for the quantity of interest in this problem, the
rate of production $\Gamma$
\begin{eqnarray}\label{eq:prodrate}
\Gamma &=& \frac{d\mathcal{P}}{dt} \to \frac{\mathcal{P}}{T} \nonumber\\
\Gamma &=& \frac{c}{\hbar} \left( \prod_{f} \int \frac{d^3
\vec{p}_{f}}{(2\pi)^3} \frac{1}{2E_{f}/c} \right)
\frac{1}{2E_{\nu}/c}\ \sum_{n=-\infty}^{\infty} (2\pi)^4
\delta^4(q_e + q_{\bar{e}} + p_{\nu'} - p_{\nu} - n k)\
\overline{\bigl| \mathcal{M}_n\bigr|^2}\,.
\end{eqnarray}

It may be most appropriate to think of the total production rate
$\Gamma$ as the sum of the production rates for all of the
individual processes involving $n$ photons
\begin{equation}\label{eq:totalpr}
\Gamma = \sum_{n = -\infty}^{\infty} \Gamma_n\,,
\end{equation}
where
\begin{equation}\label{eq:prn}
\Gamma_n = \frac{c}{\hbar} \left( \prod_{f} \int \frac{d^3
\vec{p}_{f}}{(2\pi)^3} \frac{1}{2E_{f}/c} \right)
\frac{1}{2E_{\nu}/c}\ (2\pi)^4\ \delta^4(q_e + q_{\bar{e}} +
p_{\nu'} - p_{\nu} - n k)\ \overline{\bigl|
\mathcal{M}_n\bigr|^2}\,.
\end{equation}

\section{Results\label{sec:results}}
Our calculation proceeds by computing the individual production
rates $\Gamma_n$ (equation~(\ref{eq:prn})) and then summing to
find the total rate $\Gamma$ (equation~(\ref{eq:totalpr})). Though
the constraint in equation~(\ref{eq:nconstraint2}) imposes a lower
bound on the summation over $n$ for the total rate, there is no
such upper bound. This means that the calculation should include
computations of the individual production rates for $n$ out to
infinity.  Fortunately, the individual production rates $\Gamma_n$
fall off exponentially for large $n$. How fast (or slowly) these
rates fall off depends on what choices are made for the incoming
photon energy $E_\gamma$ and the magnitude of the vector potential
$a$. For the cases in which the individual rates fall off slowly,
we calculate the rates to sufficiently large $n$ such that we can
characterize the nature of the exponential decay. Using this
characterization to fit the rates at large $n$ allows us to
approximate the summation.

We choose initial conditions that closely approximate today's
high-intensity laser system.  These conditions are based on the
Terawatt High-intensity Optical Research (THOR) laser at the High
Intensity Laser Science Group at the University of Texas at
Austin. The THOR laser is a $20\,\mathrm{TW}$ laser beam centered
around a wavelength of $800\,\mathrm{nm}$ with an electric field
strength $|\vec{E}|$ on the order of $10^{10} \sim
10^{11}\,\mathrm{V/cm}$ \cite{Ditmire}.  For our calculations we
choose an electric field strength of $|\vec{E}| = 5 \cdot
10^{10}\,\mathrm{V/cm}$.  The choice of wavelength and electric
field strength uniquely specify the electromagnetic field
parameters in our problem.  Using the customary relationships, the
photon energy is
$$
E_\gamma = hc/\lambda\,,
$$
and the magnitude of the vector potential is
$$
a = |\vec{E}| \hbar / E_\gamma\,.
$$

\begin{figure*}
\centering \includegraphics{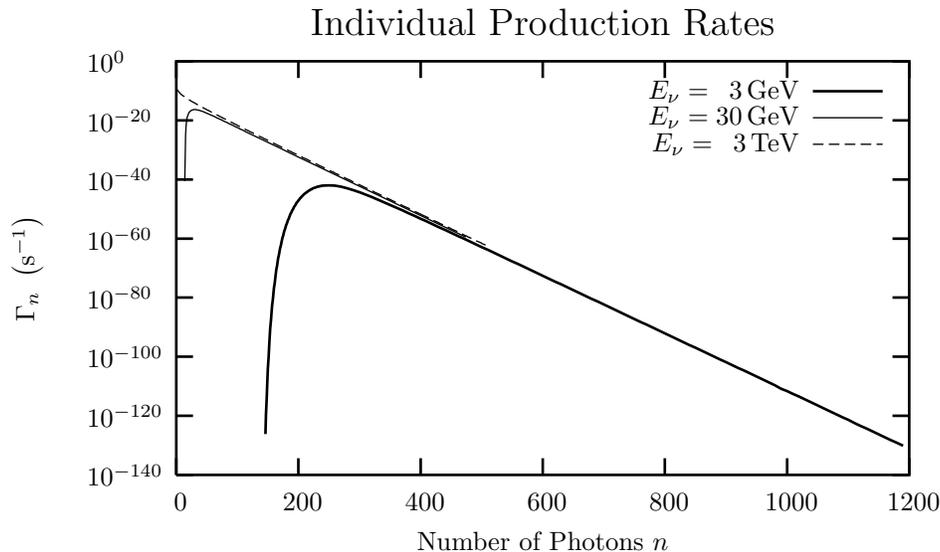} \caption{The profile of the
individual production rates $\Gamma_n$ as a function of $n$ for
incoming neutrino energies of 3\,GeV, 30\,GeV, and
3\,TeV.\label{fig:fig1}}
\end{figure*}

The profile of the individual production rates $\Gamma_n$ as a
function of the number of photons $n$ is given in
Fig.~\ref{fig:fig1} for incoming neutrino energies of 3\,GeV,
30\,GeV, and 3\,TeV.  Notice that for a given choice of photon
energy and vector potential magnitude, the profiles shown in
Fig.~\ref{fig:fig1} all follow the same exponential decay.

The individual rates that make up these profiles are summed in
order to find the total production rate
(equation~(\ref{eq:totalpr})). We present the results of this
summation in Table~\ref{tab:drates} for a range of neutrino
energies from 1\,GeV to 300\,TeV.  The total rate of production is
also presented for a laser field that is able to generate the same
field strengths but with a much smaller wavelength of light
$\lambda = 100\,\mathrm{nm}$.  We note that the difference in
wavelengths has the most significant effect at low incoming
neutrino energies.

\begin{table}
\centering \begin{tabular}{|r||l|l|}
\multicolumn{3}{c}{}\\\hline%
\multicolumn{1}{|c||}{$E_\nu$} & \multicolumn{2}{c|}{$\Gamma$
$(\mathrm{s^{-1}})$}\\\cline{2-3}
\hspace{1.5cm} & \ $\lambda= 800\,\mathrm{nm}$\ & \ $\lambda= 100\,\mathrm{nm}$\ \\\hline%
\ $1\,\mathrm{GeV}$   &\ $9.1 \cdot 10^{-93}$ &\ $6.7 \cdot 10^{-58}$\\
\ $3\,\mathrm{GeV}$   &\ $3.3 \cdot 10^{-41}$ &\ $8.1 \cdot 10^{-29}$\\
\ $10\,\mathrm{GeV}$  &\ $3.8 \cdot 10^{-22}$ &\ $1.1 \cdot 10^{-17}$\\
\ $30\,\mathrm{GeV}$  &\ $6.7 \cdot 10^{-16}$ &\ $8.3 \cdot 10^{-14}$\\
\ $100\,\mathrm{GeV}$ &\ $5.5 \cdot 10^{-13}$ &\ $8.1 \cdot 10^{-12}$\\
\ $300\,\mathrm{GeV}$ &\ $1.3 \cdot 10^{-11}$ &\ $5.7 \cdot 10^{-11}$\\
\ $1\,\mathrm{TeV}$   &\ $1.2 \cdot 10^{-10}$ &\ $3.0 \cdot 10^{-10}$\\
\ $3\,\mathrm{TeV}$   &\ $6.0 \cdot 10^{-10}$ &\ $1.2 \cdot 10^{-9}$\\
\ $10\,\mathrm{TeV}$  &\ $2.9 \cdot 10^{-9}$  &\ $5.1 \cdot 10^{-9}$\\
\ $30\,\mathrm{TeV}$  &\ $1.1 \cdot 10^{-8}$  &\ $1.8 \cdot 10^{-8}$\\
\ $100\,\mathrm{TeV}$ &\ $4.8 \cdot 10^{-8}$  &\ $7.1 \cdot 10^{-8}$\\
\ $300\,\mathrm{TeV}$ &\ $1.7 \cdot 10^{-7}$  &\ $2.4 \cdot
10^{-7}$\\\hline
\end{tabular}
\caption{The total production rate $\Gamma$ for a given incoming
neutrino energy $E_\nu$ in the presence of an laser field of
electric field strength $|\vec{E}| = 5 \cdot
10^{10}\,\mathrm{V/cm}$ and wavelength
$\lambda$.\label{tab:drates}}
\end{table}

\begin{figure*}
\centering \includegraphics{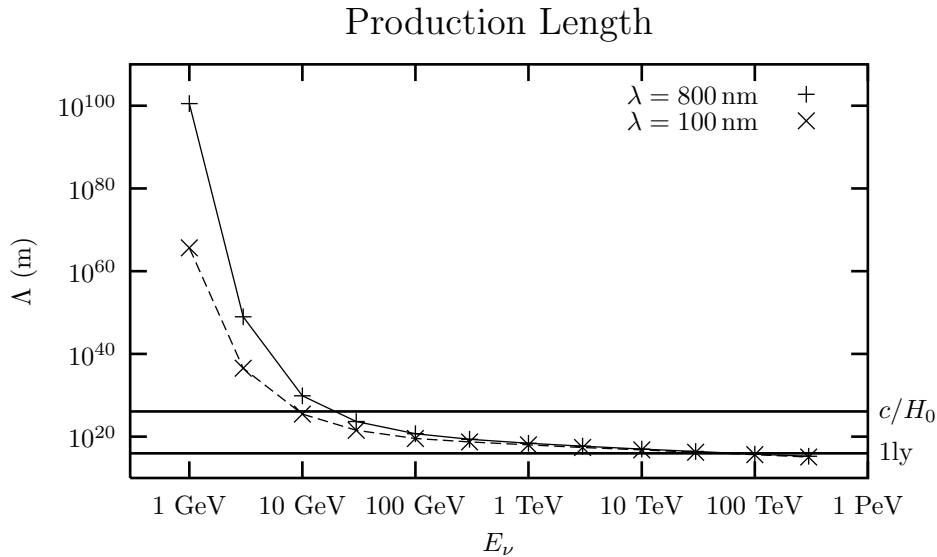} \caption{The production
lengths $\Lambda$ as a function of incoming neutrino energy
$E_\nu$ for a laser field of wavelengths 100\,nm and 800\,nm at an
electric field strength of $|\vec{E}| = 5 \cdot
10^{10}\,\mathrm{V/cm}$.  The dark lines correspond to the Hubble
length $c/H_0 \sim 1.2 \cdot 10^{26}\,\mathrm{m}$ and one light
year $1\,\mathrm{ly} = 9.46 \cdot
10^{15}\,\mathrm{m}$.\label{fig:fig2}}
\end{figure*}

Rather than considering the total rate of production for this
process, it may have more physical significance if we consider the
production length $\Lambda$.  The production length is the
distance that a neutrino must travel in the laser field such that
its likelihood of producing an electron-positron pair is $1 - 1/e
\simeq 63\%$.  That is, the probability for production is
$$
\mathcal{P} = 1 - e^{-\ell/\Lambda}\,,
$$
where $\ell$ is the distance traveled in the field, and the
production length is simply $\Lambda = c / \Gamma$.  The
production lengths for the rates tabulated in
Table~\ref{tab:drates} are shown in Fig.~\ref{fig:fig2}.

From inspection of Fig.~\ref{fig:fig2}, one can see how feeble
this interaction is. For the production length to drop below the
Hubble length, an estimate for the size of the universe, the
incoming neutrino energy must exceed 10\,GeV.  And at energies
near 1\,PeV the production length is still on the order of a light
year.

To further stress the small likelihood of pair production under
these conditions, we can consider this interaction at a neutrino
source such as the Neutrinos at the Main Injector (NuMI) facility
at Fermilab.  If we assume the most ideal set of specifications
for the neutrino beam \cite{Kopp:2002iu} and simplify the beam to
be uniform, we estimate the neutrino flux to be on the order of
$10^{19}/\mathrm{m^2}/\mathrm{s}$~per~pulse with a pulse every
2\,s.  The energy profile of the beam is nontrivial.  There is a
low-energy component peaked at 3\,GeV, a medium-energy component
peaked at 7\,GeV, and a high-energy component peaked at 15\,GeV,
but the beam will contain neutrinos with energy up to 120\,GeV
\cite{Zwaska}.  However, for estimation purposes, we simply assume
that all of the neutrinos are at 15\,GeV and are uniformly
distributed throughout the pulse.  Using these conditions for the
NuMI beam, realistic conditions for the laser beam, and assuming a
detection region about 1\,m long, we estimate that there would be
an event once every $10^{22}\,\mathrm{yr}$.

Given that our estimate of the time scale for detection of events
is twelve orders of magnitude greater than the age of the
universe, future work must turn towards astronomical and
cosmological sources of neutrinos and fields.  In astronomical
phenomena where the magnetic field strengths can reach very high
magnitudes, there exists the possibility of significant
stimulation of this interaction.

\begin{acknowledgments}
I am especially grateful to D.~Dicus for his advice and
suggestions throughout this work.  I also wish to extend special
thanks to T.~Ditmire and G.~Dyer for the useful information on
ultrashort pulsed lasers, and to R.~Zwaska for the beneficial
conversations concerning the NuMI/MINOS experiment.
\end{acknowledgments}

\appendix*\section{The Scattering Amplitude
{}$\mathbf{\mathcal{M}_n}$\label{ap:M}}%
As pointed out in the text, the scattering amplitudes
$\mathcal{M}_n$ are related to the $S$-matrix by
equation~(\ref{eq:Smatrix3})
$$
S = \imath (2\pi)^4 \sum_{n=-\infty}^{\infty} \delta^4(q_e +
q_{\bar{e}} + p_{\nu'} - p_{\nu} - n k)\ \mathcal{M}_n\,.
$$
The scattering amplitudes are determined to be
\begin{widetext}
\begin{eqnarray}\label{eq:aMplitude}
\mathcal{M}_n &=&  - \frac{4\pi \alpha}{2^2 \cos^2
\theta_{\mathrm{W}} \sin^2\theta_{\mathrm{W}}}\
\overline{u}_{s_{\nu'}}(p_{\nu'}) \gamma^\sigma \left(\frac{1}{2}
- \frac{1}{2} \gamma^5\right) u_{s_\nu}(p_{\nu})\
\frac{g_{\mu\sigma} - (p_{\nu}-p_{\nu'})_\mu
(p_{\nu}-p_{\nu'})_\sigma/(m_Z c)^2}{(p_{\nu}-p_{\nu'})^2 - (m_Z
c)^2 + \imath \Gamma_Z m_Z c^2} \nonumber\\
&& \times\ \bar{u}_{s_e}(p_e) \bigl[{C_0}_n(\zeta,\delta)
\Gamma_0^\mu + {C_x}_n(\zeta,\delta) \Gamma_x^\mu +
{C_y}_n(\zeta,\delta) \Gamma_y^\mu 
+ {C_{xx}}_n(\zeta,\delta) \Gamma_{xx}^\mu +
{C_{xy}}_n(\zeta,\delta) \Gamma_{xy}^\mu +
{C_{yy}}_n(\zeta,\delta) \Gamma_{yy}^\mu \bigr]
v_{s_{\bar{e}}}(p_{\bar{e}})\,. \nonumber\\
&&
\end{eqnarray}
\end{widetext}
The $C_n(\zeta,\delta)$'s are coefficients that depend on the
Bessel functions
\begin{subequations}\label{eq:Coef}
\begin{eqnarray}
{C_0}_n(\zeta,\delta) &=& J_n(\zeta) e^{\imath n \delta}
\label{eq:C0}\\
{C_x}_n(\zeta,\delta) &=& \frac{1}{2} \left( J_{n+1}(\zeta)
e^{\imath (n+1) \delta} + J_{n-1}(\zeta) e^{\imath (n-1) \delta}
\right) \label{eq:Cx} \\
{C_y}_n(\zeta,\delta) &=& \frac{1}{2\imath} \left( J_{n+1}(\zeta)
e^{\imath (n+1) \delta} - J_{n-1}(\zeta) e^{\imath (n-1) \delta}
\right) \label{eq:Cy} \\
{C_{xx}}_n(\zeta,\delta) &=& \frac{1}{2} J_n(\zeta) e^{\imath n
\delta} + \frac{1}{4} \left( J_{n+2}(\zeta) e^{\imath (n+2)
\delta} + J_{n-2}(\zeta) e^{\imath (n-2) \delta} \right)
\label{eq:Cxx} \\
{C_{xy}}_n(\zeta,\delta) &=& \frac{1}{4\imath} \left(
J_{n+2}(\zeta) e^{\imath (n+2) \delta} - J_{n-2}(\zeta) e^{\imath
(n-2) \delta} \right) \label{eq:Cxy} \\
{C_{yy}}_n(\zeta,\delta) &=& \frac{1}{2} J_n(\zeta) e^{\imath n
\delta}  - \frac{1}{4} \left( J_{n+2}(\zeta) e^{\imath (n+2)
\delta} + J_{n-2}(\zeta) e^{\imath (n-2) \delta} \right)\,.
\label{eq:Cyy}
\end{eqnarray}
\end{subequations}
Recall that $\zeta$ and $\delta$ are defined through
equations~(\ref{eq:zddef}).  The $\Gamma_i^\mu$ matrices are given
by
\begin{subequations}\label{eq:gammas}
\begin{eqnarray}
\Gamma_0^\mu &=& \gamma^\mu \left(-\frac{1}{2}+2
\sin^2\theta_{\mathrm{W}} + \frac{1}{2} \gamma^5\right)
\label{eq:G0} \\
\Gamma_x^\mu &=& -\frac{ea \gamma^1 \nk}{2 p_e\cdot k}
\Gamma_0^\mu - \Gamma_0^\mu \frac{ea \gamma^1 \nk}{2 p_{\bar{e}}
\cdot k} \label{eq:Gx} \\
\Gamma_{y}^\mu &=& -\frac{ea \gamma^2 \nk}{2 p_e\cdot k}
\Gamma_0^\mu - \Gamma_0^\mu \frac{ea \gamma^2 \nk}{2 p_{\bar{e}}
\cdot k} \label{eq:Gy} \\
\Gamma_{xx}^\mu &=& \frac{ea \gamma^1 \nk}{2 p_e\cdot k}
\Gamma_0^\mu \frac{ea \gamma^1 \nk}{2 p_{\bar{e}} \cdot k}
\label{eq:Gxx} \\
\Gamma_{xx}^\mu &=& \frac{ea \gamma^1 \nk}{2 p_e\cdot k}
\Gamma_0^\mu \frac{ea \gamma^2 \nk}{2 p_{\bar{e}} \cdot k} +
\frac{ea \gamma^2 \nk}{2 p_e\cdot k} \Gamma_0^\mu \frac{ea
\gamma^1 \nk}{2 p_{\bar{e}} \cdot k} \label{eq:Gxy} \\
\Gamma_{yy}^\mu &=& \frac{ea \gamma^2 \nk}{2 p_e\cdot k}
\Gamma_0^\mu \frac{ea \gamma^2 \nk}{2 p_{\bar{e}} \cdot k}\,,
\label{eq:Gyy}
\end{eqnarray}
\end{subequations}
where $\gamma^1$ and $\gamma^2$ are Dirac matrices.

\end{document}